# New designing of E-Learning systems with using network learning

Amin Daneshmand Malayeri
Department of Computer Engineering
Young Researchers Club, Malayer Azad University
Malayer, Iran
amin.daneshmand@gmail.com

Jalal Abdollahi
Department of Computer Engineering
Hamedan University of Technology
Hamedan , Iran
jalal.abdollahi66@gmail.com

*Abstract*—One of the most applied learning in virtual spaces is using E-Learning systems. Some E-Learning methodologies has been introduced, but the main subject is the most positive feedback from E-Learning systems. In this paper, we introduce a new methodology of E-Learning systems entitle " Network Learning" with review of another aspects of E-Learning systems. Also, we present benefits and advantages of using these systems in educating and fast learning programs.
Network Learning can be programmable for every education system and it is flexible with too positive results.

## I. Introduction

Online learning is increasingly becoming the norm and part of the mainstream in higher education. As universities and community colleges across the U.S. continue to offer a growing number of online courses, interest in the field of distance learning and distance learners is growing. Much research is devoted to learner assessments, solid instructional practices as well as learner outcomes and satisfaction. One important avenue for learning about learner needs is to look at the feedback students themselves provide about successful and effective e-learning. The inquiry was driven by the question: what are community college learners' needs in online environments as explained by the students? The author considers it crucial to understand learner needs and provide means via course design and instruction to meet them. The underlying assumption is that a learner-centered online program will foster meaningful learning for its participants [1].

Looking at the efficacy of student learning based on learner feedback in online learning is particularly difficult because much of the research done in the late 1990s and early 2000s is somewhat antiquated [2]. Technology has been changing so fast that many of the obstacles and reasons for dissatisfaction reported by earlier studies have been rendered obsolete by advances in technology.

Course management systems (CMS) today provide a level of interaction unimaginable a few years ago. Further successful analysis of undergraduate and community college students' experiences in online setting is hindered by the scarcity of available data[3].

Interpretation of available data is further complicated by the multi-cultural background and diversity of researchers and research subjects. Looking at a wealth of data representing diverse cultural backgrounds is usually desirable; however, in the context of student satisfaction in e-learning, cultural diversity must be considered with caution. There is a strong possibility that e-learners in Taiwan or Hong Kong have a different idea about learner needs, success, and satisfaction than e-learners in the American Southwest who attend a local community college. Student perceptions of what constitute learner needs and learner outcomes are colored by one's culture. Further inquiry is required to establish the extent to which research conducted on other continents of the world is applicable to non-traditional American students. As a result, the current inquiry initially reviewed but did not incorporate the findings of studies where: a) the subjects were graduate students or beyond their bachelor's degree and b) the subjects came from an educational system that is markedly different from the American educational system in its student –teacher interaction and communication. After a careful review of the available data, the author decided to incorporate a handful of research papers that reported on student feedback of American graduate students because there were not enough studies done with solely undergraduate and community college students[3,7].

Software and learning platforms are a key issue in attempting to develop sustainable strategies for the use of ICT in education. However, there are some difficulties in developing such strategies, because the very different learning applications used in different contexts and institutions. In addition to software designed specifically or learning, many courses will present in work-based applications , spreadsheets, statistical applications. Almost all educational institutions use general office, financial and management applications. Many may also use specialist management information systems.

Sustainability in this respect takes on a number of different perspectives. The obvious issue is the cost of licensing software. A second issue is software support. A third is allowing learners access to modern applications, in terms of functionality, reliability and attractiveness. Form a work based learning perspective it is also important that learners are able to use software and applications they ill come across in the workplace. However, the fast speed of software devel-opment, the endless version releases and the expense of upgrades is a considerable barrier for many institutions.
In this section I will explore three issues. The first is Learning Management Systems (LMS). The second is the potential of Open Source Software for education and training. The third is new partnerships for developing software and architectures [4].

## II. SOME E-LEARNING STRATEGIES

### A) Sharing resources

With the present cost of hardware, resources will seldom be sufficient for most public education and training providers. One answer lies in the sharing and more intensive use of scarce resources. Computers used in schools in the daytime can be used for community education in the evening. Design schools can allow use of high performance machines for project work by schools. Access points can be set up in local housing facilities. Greater co-ordination between library services and schools can allow access to scarce bandwidth. Universities can provide and maintain networks for schools and community education providers. Ideally, these initiatives need to be coordinated on a local basis [5].

### B) Using older technology

Whilst most accounting systems depreciate hardware over a period of three years, there is nothing to say that machines may not have a much longer useful lifespan. I recently ran a workshop in a computer training centre in a community centre on a large and poor housing complex on the outskirts of Dublin where not one of the computers was purchased after 1997. Although perhaps due to the excellence of the open source software used in the session, not one machine crashed during a whole days training. Modern computers are generally over-specified for the needs of most e-learning applications. Servers do not require a high performance machine. Many countries have schemes where older computers are recycled from industry and commerce to education.

Of course older machines tend to require more maintenance and anecdotal evidence suggests this can be a major problem for many schools. Students and parents seem to be frequently involved in this work. In this respect sustainability may lie in a greater community involvement and orientation. Local companies and computer professionals can contribute resources and expertise to designing, configuring and maintaining hardware and software networks.

Older computers may not run more modern and higher memory applications. It is doubtful how necessary these are for most subjects and learning contexts. Most open source applications are designed to run on lower specification machines and earlier operating systems. The Linux operating system will run on most computers.

Voice over IP (VOIP) applications which carry audio data through the web may offer schools and other educational institutions considerable cost savings in the future, savings which could be re-invested in hardware and infrastructure.

### C) Mobile devices

One possible answer for access to hardware which is exciting some interest is the possible use of handheld devices, palmtop computers, PDAs and mobile telephones. Whilst many are skeptical due to the small screen size, other researchers have pointed to the intensive use of mobile telephones by young people for a wide range of applications. Furthermore, there is widespread interest in lesser-developed countries, where the cost of hardware is prohibitive to introducing e-learning [6].

These methods are available in some E-learning sites, but we offer using Network Learning as describe in III.

## III. NETWORK LEARNING SYSTEMS

This system can be defined in two state of programming. The first state is like a forum for transferring of scientific experiences of users. By this system we make a cluster learning space for all of users without any limit for them as figure 1.

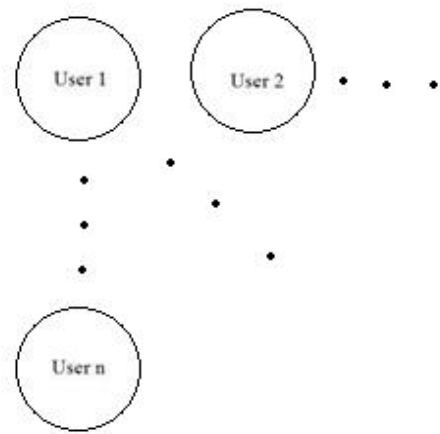

Figure 1. Cluster network learning for all of users

By using this method for "n" users number of relations between users (UR) is :

$$UR = \frac{n(n-1)}{2} \qquad (1)$$

The second state is priority-based for introducing level of availability. At this state, every user can use from network documents by his/her activity level. Users that are active, can take more documents for their educational goals than inactive users. Growth Rate of this network is up to users activity. We can define a Pre-Growth Rate (PGR) for these E-learning networks as (2):

$$PGR = KnT \qquad (2)$$

That "K" is a coefficient for showing activity rate in a period of time (T).

Educational feedback in this state is a value with depending on users activity. We can label to users by their activity. Uploading / Downloading level for documents can be define for users by their Growth Rate. This method can encourage all of users for using educational documents faster than past. This is one of the best way for performing "fast learning" methodology. Also, we can control level of availability of users by measuring their Growth Rate.

## IV. LEARNING DOCUMENTS PUBLISHING BY NETWORK LEARNING

Network Learning can make an open learning space. All of users in this system can share their documents together. We can define open schools, colleges and universities by this methodology for founding a new base of Learning method entitle "Open Learning ". By making a database for educational background and feedback and publishing them as some experiences, performing of this methodology can be occurred. Especially, we can define intricacy learning networks to covering all of virtual learning systems.

This methodology has been shown in Figure 2.

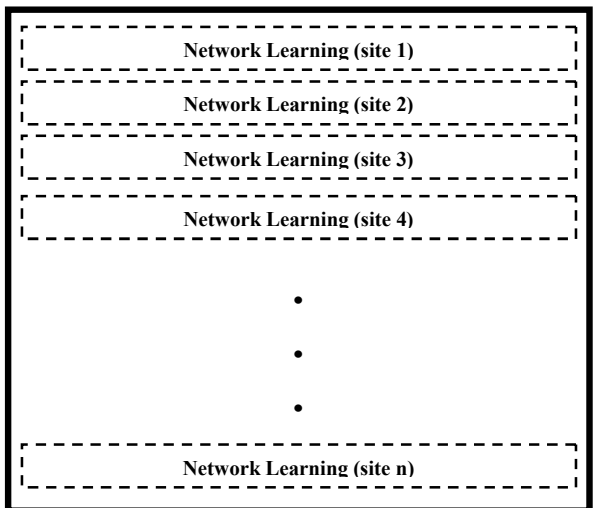

Figure 2. intricacy network learning

## V. CONCLUSION

In this paper, we introduce some new methodologies in E-Learning with base of Network Learning. Nowadays, some of students and even teachers are interested in making a virtual and semantic space for their educational goals and receiving positive feedback.

Designing E-schools, E-colleges and E-universities by using network learning is easier and with lower cost in comparison to other solutions. Even, we can design a space for encouraging students for more trying to scientific documents transferring and publishing them. Attention to internet and depended equipments and infrastructures can make a better design for these kind of web-based spaces and its software.